\documentclass{elsart}
\usepackage{graphicx}
\usepackage{color}
\usepackage{amssymb}
\begin{document}
\begin{frontmatter}
\title{Backward-angle photoproduction of $\pi^0$ mesons on the proton 
at \mbox{\boldmath $E_\gamma = $}1.5--2.4~GeV}
\author[rcnp]{M.~Sumihama},
\ead{sumihama@rcnp.osaka-u.ac.jp}
\author[pusan]{J.K.~Ahn},
\author[konan]{H.~Akimune},
\author[jasri]{Y. Asano},
\author[taiwan]{W.C.~Chang},
\author[jasri]{S.~Dat\'{e}},
\author[jasri]{H. Ejiri},
\author[kyoto]{H.~Fujimura}, 
\author[rcnp,jaeri]{M.~Fujiwara},
\author[ohio]{K.~Hicks},
\author[rcnp]{T.~Hotta}, 
\author[kyoto]{K.~Imai},
\author[lns]{T.~Ishikawa},
\author[yamagata]{T.~Iwata},
\author[chiba]{H.~Kawai},
\author[seoul]{Z.Y. Kim},
\author[rcnp]{K.~Kino},
\author[rcnp]{H.~Kohri},
\author[jasri]{N. Kumagai},
\author[wakayama]{S. Makino},
\author[rcnp]{T.~Matsumura\thanksref{now-nda}},
\author[rcnp]{N.~Matsuoka},
\author[rcnp]{T.~Mibe\thanksref{now-jlab}},
\author[kyoto]{M.~Miyabe},
\author[TIT]{Y. Miyachi},
\author[rcnp]{M.~Morita},
\author[rcnp]{N.~Muramatsu},
\author[rcnp]{T.~Nakano},
\author[kyoto]{M.~Niiyama},
\author[osaka]{M.~Nomachi},
\author[jasri]{Y.~Ohashi},
\author[jasri]{H. Ohkuma},
\author[chiba]{T.~Ooba},
\author[taiwan]{D.S.~Oshuev},
\author[canada]{C.~Rangacharyulu},
\author[osaka]{A.~Sakaguchi},
\author[osaka]{T.~Sato},
\author[minnesota]{P.M.~Shagin},
\author[chiba]{Y.~Shiino},
\author[lns]{H.~Shimizu},
\author[osaka]{Y.~Sugaya},
\author[jasri]{H.~Toyokawa},
\author[akita]{A. Wakai},
\author[taiwan]{C.W.~Wang},
\author[taiwan]{S.C.~Wang},
\author[ILI]{K. Yonehara},
\author[rcnp]{T.~Yorita},
\author[rcnp,kyoto]{M.~Yosoi},
\author[msu]{R.G.T.~Zegers}

\address[rcnp]{
\it Research Center for Nuclear Physics, Osaka University,
Ibaraki~567-0047, Japan} 
\address[jaeri]{
\it Kansai Photon Science Institute, Japan Atomic Energy Agency, 619-0215 
Kizu, Japan} 
\address[pusan]{
\it Department of Physics, Pusan National University, Busan~609-735, Korea}
\address[konan]{
\it Department of Physics, Konan University, Kobe~658-8501, Japan}
\address[taiwan]{
\it Institute of Physics, Academia Sinica, Taipei~11529, Taiwan}
\address[jasri]{
\it Japan Synchrotron Radiation Research Institute,
Sayo, Hyogo~679-5198, Japan} 
\address[kyoto]{
\it Department of Physics, Kyoto University, Kyoto~606-8502, Japan}
\address[ohio]{
\it Department of Physics and Astronomy, Ohio University,
Athens, Ohio~45701, USA} 
\address[seoul]{
\it School of Physics, Seoul National University, Seoul, 151-747, Korea}
\address[yamagata]{
\it Department of Physics, Yamagata University, Yamagata 990-8560, Japan}
\address[wakayama]{
\it Wakayama Medical College, Wakayama 641-8509, Japan}
\address[TIT]{
\it Department of Physics, Tokyo Institute of Technology, Tokyo 152-8551, 
Japan} 
\address[chiba]{
\it Graduate School of Science and Technology, Chiba University,
Chiba~263-8522, Japan} 
\address[canada]{
\it Department of Physics and Engineering Physics , University of Saskatchewan,
Saskatoon, Saskatchewan~S7N5E2, Canada} 
\address[minnesota]{
\it School of Physics and Astronomy, University of Minnesota, Minneapolis, Minnesota 55455}
\address[akita]{
\it Akita Research Institute of Brain and Blood Vessels, Akita 010-0874, Japan}
\address[osaka]{
\it Department of Physics, Osaka University, Toyonaka~560-0043, Japan} 
\address[lns]{
\it Laboratory of Nuclear Science, Tohoku University,
Sendai~982-0826, Japan}
\address[ILI]{
\it Illinois Institute of Technology, Chicago, Illinois 60616, USA}
\address[msu]{National Superconducting Cyclotron Laboratory,
Michigan State University, Michigan 48824, USA}

\thanks[now-nda]{
Present address: Department of Applied Physics,
National Defense Academy, Yokosuka~239-8686, Japan}
\thanks[now-jlab]{
Present address: Department of Physics and Astronomy, Ohio University, Ohio 45701, USA}

\begin{abstract}
Differential cross sections and photon beam asymmetries for $\pi^0$ 
photoproduction have been measured at $E_\gamma$ = 1.5--2.4~GeV 
and at the $\pi^0$ scattering angles, --1 $<$ cos$\Theta_{c.m.} <$ --0.6. 
% in the center of mass system at the SPring-8/LEPS facility. 
%The differential cross sections have been obtained with sufficient statistics 
The energy-dependent slope of differential cross sections for $u$-channel 
$\pi^0$ production has been determined. An enhancement at backward angles 
is found above $E_\gamma$ = 
2.0~GeV. This is inferred to be due to the $u$-channel contribution and/or 
resonances. Photon beam asymmetries have been obtained for the first time at backward angles. A strong angular 
dependence has been found at $E_\gamma >$ 2.0~GeV, which may be due to 
the unknown high-mass resonances. 
\end{abstract}

\begin{keyword}
\PACS 13.60.Le \sep 14.40.Aq \sep 25.20.Lj
\end{keyword}
\end{frontmatter}

% Introduction

% MAID/SAID other models
%Missing resonances
%Scaling, transition region

Pion photoproduction has been well studied both experimentally and 
theoretically in the spectroscopy of N* and $\Delta$* resonances. 
Many baryon resonances were found and their characteristics were 
determined at the total energy, $W <$ 1.7~GeV~\cite{PDG}. 
% (up to P$_{13}$(1720))
However 
%baryon resonances are not well established above 
%1.7 GeV. 
there are many higher-mass resonances (one or two-star 
resonances) which are not well established and are called missing 
resonances~\cite{PDG,Capstick}. 
Constituent quark models predict more baryon resonances than those 
observed experimentally~\cite{Capstick}. 
Identification of these missing resonances is important to 
understand the quark-gluon structure of a nucleon. 
%Some of baryon resonances decouple from the pion-nucleon channel so 
%that the other channels are more sensitive to those resonances.  
%In fact, the $K^+$, $\eta$ and $\omega$ photoproductions are 
%discussed connecting with missing resonances~\cite{kaon,eta,omega}. 

There is a possibility to obtain new information on baryon resonances 
in pion photoproduction by measuring polarization observables at high 
energies. 
Some weakly excited resonances are obscured due to other strong resonances 
which have large decay widths, making it difficult to demonstrate their 
existance only from the cross section data. 
%It is suggested that 
Polarization observables are good means to extract the missing  
resonances~\cite{Dutta-pi0}. In fact, a strong angular 
dependence was found in the induced polarization measured using circularly 
polarized photons for $\pi^0$ photoproduction at JLab, suggesting 
the possibility of a relatively high spin resonance~\cite{halla-pi0}. 

Photo- and electro-production of mesons have been studied at JLab, ELSA, 
GRAAL, and SPring-8/LEPS via the measurements of not only cross sections 
but also polarization data at multi-GeV energies. Theoretical models of these 
meson production processes can be applied to higher energies to investigate 
the production mechanism and higher mass resonances. A partial wave analysis 
SAID gives a good fit to the experimental data up to $W$ = 2.15~GeV, 
and the characteristics of baryon resonances are determined~\cite{said}. 
A unitary isobar model, MAID, has been improved and extended up to $W$ = 
2.0~GeV by including 13 four-star resonances up to $F_{37}$(1950)~\cite{maid}. 
Dynamical coupled-channel analyses have been performed by combining the 
data of different reaction modes~\cite{SL,EBAC,Bonn}. 
%The analyses play an important rule in order to extract N* parameters.  
%The results of calculations should have the same masses, widths and couplings 
%of baryon resonances in all reaction modes. 
The presence of higher mass resonances is shown in the analysis by combining the photoproduction data on N$\pi$, N$\eta$ and 
$K$Y channels~\cite{EBAC,Bonn}.   
However, there are still model uncertainties, especially for relatively 
higher-mass resonances. 
% and no model is well established above W = 2.0 GeV. 
The model uncertainties come in part from fewer data points at higher 
energies. 
It is important to measure more precise differential cross sections and 
polarization data around $W$ = 2.0~GeV in a full angular coverage. 
%The precise measurements of polarization observables give a test of model 
%assumptions. 

%introduction for LEPS data 
%The differential cross sections and photon beam asymmetries of the $\gamma p 
%\rightarrow \pi^0 p$ reaction have been measured using linearly 
%polarized photons with the photon energies 1.5 GeV $<$ $E_\gamma <$ 2.4 GeV 
%and at --1 $<$ cos$\Theta_{c.m.} <$ --0.6 of the $\pi^0$ scattering angle. 
The LEPS photon energy, $E_\gamma$ = 1.5--2.4~GeV, sits in a 
transition region from nucleon-meson degrees of freedom to quark-gluon 
degrees of freedom. 
The cross sections are known to approximately follow the constituent counting 
rules above the resonance region and at large scattering angles~\cite{scal}. 
The scaling behavior of the differential cross sections has been observed for 
many photoproduction measurements~\cite{scalexppi,scalexp}. Although the LEPS 
photon energy is slightly lower than the energy where the scaling behavior 
appears (the onset of scaling), the LEPS data provide information on the early 
onset of scaling and quark-hadron duality in the transition 
region~\cite{halla-pi0,duality}. 

At very backward angles, the production mechanism is expected to be affected 
mainly by the $u$-channel contribution where a proton or nucleon resonance is 
the exchanged baryon. 
%The higher mass N* and $\Delta$* resonances likely have smaller amplitudes 
%than a nucleon due to their heavy masses. 
The differential cross section data of meson 
photoproduction show the behavior of the nucleon Regge pole $s^{2\alpha(u)-2}$ 
at high energies $E_\gamma >$ 3~GeV ~\cite{OmegaInU,Guidal}. The LEPS data provide 
information on the $u$-channel mechanism at $E_\gamma = $1.5--2.4~GeV and 
at the $u$-channel kinematics where there is no experimental data of $\pi^0$ 
photoproduction. 

% Experimental setup
The experiment was carried out at the Laser-Electron-Photon beam line 
of the Super Photon ring 8-GeV facility (SPring-8/LEPS)~\cite{NAK01}. 
A multi-GeV photon beam was produced by backward-Compton scattering (BCS)  
from the head-on collision between Ar-ion laser photons with a 351-nm wave 
length and the circulating 8-GeV electrons in the storage ring. The linearly 
polarized photon beam was obtained from the BCS process with 
linearly-polarized laser photons. The polarization of the photon 
beam was about 95\% at the maximum energy, 2.4~GeV and about 55\% at the 
lowest energy, 1.5~GeV. 
The photon beam energy was determined by measuring the energy of 
the recoil electron from Compton scattering with a tagging counter 
which consisted of 2 layers of a combination of a hodoscope and a silicon 
strip detector.    
The photon energy resolution was 15 MeV (RMS). The typical photon intensity, 
integrated from 1.5~GeV to 2.4~GeV, was 5$\times 10^5$/s. 
Half of the data were taken with vertically polarized photons 
and the other half with horizontally polarized photons. 
The polarization was switched about every 6 hours to reduce  
systematic errors in measuring of the photon beam asymmetries. 
A liquid hydrogen target with a thickness of 5.6 cm was used.  
The data were accumulated with 2.1$\times 10^{12}$ photons at the 
target in total. 

Produced protons were detected by the LEPS spectrometer covering 
forward angles.  A plastic-scintillation charge-veto counter was located 
before the target to eliminate charged particles produced by the exit 
windows and the residual gas in the vacuum pipe for the photon beam.  
The spectrometer consisted of a plastic-scintillation start counter (SC), 
a silica-aerogel 
\v{C}erenkov counter (AC), a silicon vertex detector (SVTX), a dipole 
magnet, three multi-wire drift chambers (DC1, DC2 and DC3), and a 
time-of-flight (TOF) wall. 
The field strength of the dipole magnet was 0.7 T at its center. 
The angular coverage of the spectrometer was about $\pm 0.4$ rad and 
$\pm 0.2$ rad in the horizontal and vertical directions, respectively. 
Charged particles produced at the target were detected by the SC located 
behind the target and it determined the trigger timing for the data 
acquisition system. The AC with a refractive index of 1.03 was used 
to reject $e^+e^-$ events, which were the main background in the measurement 
of photon induced hadronic reactions, at the trigger level. 
The event trigger was made by signals from the tagging hodoscopes, 
the SC and the TOF wall. 
Signals from the charge-veto counter and the AC were used as vetos. 
The typical trigger rate was 20 Hz. The dead time of the data 
acquisition system was about 3\%.  

% Analysis

Charged particles were momentum analyzed by using information from the 
silicon vertex detector and the three drift chambers. Tracks fitted within 
a 98\% confidence level were accepted for further analysis. The stop signal 
for the time-of-flight measurement was provided from signals of 40 plastic 
scintillators in the TOF wall. The start signal was provided by the RF 
signal from the 8-GeV 
electron storage ring where electrons were bunched at a minimum of 2 ns 
intervals with a width of 12 ps (RMS). The time-of-flight resolution was 180 
ps for a typical flight path length of 4.2 m. The particle mass was 
determined using the momentum, the path length and the time-of-flight. 
\begin{figure*}[h]
\begin{center}
\includegraphics*[scale=1.06]{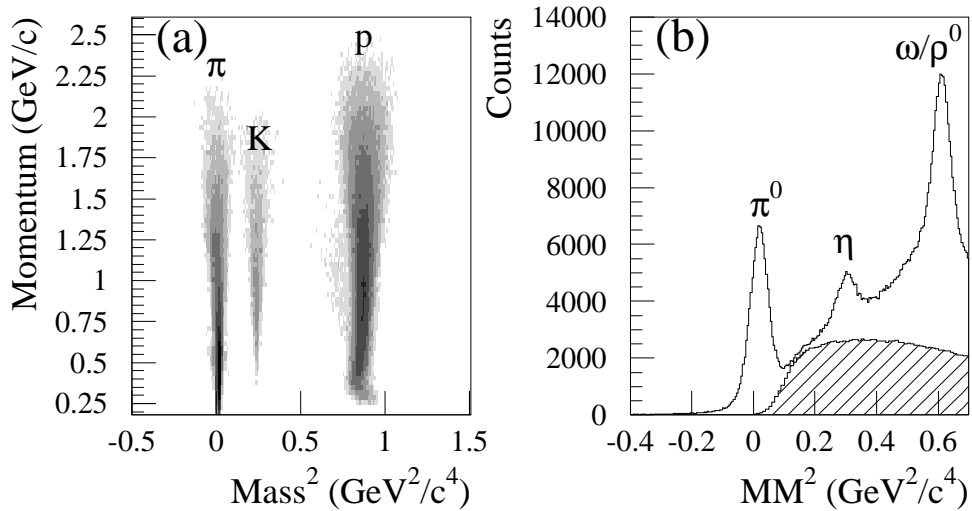}
\caption{(a) Two-dimensional scatter plot of the measured events as a 
function of momentum vs. mass squared (Mass$^2$). (b) Missing mass square 
(MM$^2$) of the $\gamma p \rightarrow p X$ reaction. 
The peaks are due to the $\pi^0$, $\eta$ and $\omega/\rho^0$ mesons. 
The hatched histogram is the distribution estimated for the double pion 
photoproduction.}
\label{MissingMass}
\end{center}
\end{figure*}
Fig.~\ref{MissingMass}(a) shows the scatter plot of the measured events as a 
function of the momentum versus the square of mass (Mass$^2$). The mass 
resolution 
depends on the momentum. The proton mass resolution was 46 MeV/c$^2$ at 2~
GeV/c momentum. Proton events were selected within 5$\sigma$ of the momentum 
dependent mass resolution. The pion and kaon contaminations were less than 
0.1\%. 

Events produced at the liquid hydrogen target (LH$_2$) were selected by their 
closest points between a track and the beam axis. Some contamination events 
stemmed from the SC behind the target. The contamination rate was large 
as the proton scattering angle was small, because the vertex resolution 
deteriorated at smaller angles. 
The contamination rate was estimated as a function of the proton 
scattering angle. The contamination rate was less than 1\% at 
cos$\Theta_{c.m.}>-0.9$, 3\% at cos$\Theta_{c.m.}=-0.925$, and 12\% at 
cos$\Theta_{c.m.}=-0.975$ in the $\pi^0$ scattering angle.  
%\begin{ratetable}[h]
%\begin{center}
%\caption{\label{tab:counts0}Number of events surviving various successive 
%event selection cuts. }
%\begin{tabular}{cc}
%Selection cuts & Events \\  \hline
%triggered events    & 1.78 $\times$ 10$^8$ \\
%one or more tracks & 4.23 $\times$ 10$^7$ \\
%single proton        & 1.42 $\times$ 10$^7$ \\
%production from LH$_2$ target & 6.28 $\times$ 10$^6$ \\
%one recoil electron in tagger & 3.78 $\times$ 10$^6$ \\ 
%\hline
%\end{tabular}
%\end{center}
%\end{table}
%Electromagnetic shower events or accidental events 
%could make a trigger, but these background events were rejected by 
%requiring one track. 
%Table~\ref{tab:counts0} shows the number of events after the selection cuts. 

The photon energy was determined by a hit position of a recoil electron 
from Compton scattering in the tagging counter. 
The missing mass (MM$^2$) of the $\gamma p \rightarrow p X$ reaction was 
shown in Fig.~\ref{MissingMass}(b). Peaks due to $\pi^0$, $\eta$ and 
$\omega/\rho^0$ photoproduction were observed.  
%The missing mass resolution depends on the momentum. The $\pi^0$'s were 
%selected within 2$\sigma$ of the momentum dependent cut. 

There are background events from the $\gamma p \rightarrow \pi\pi p$ reaction 
in the $\pi^0$ peak region. 
%When the 2$\sigma$ cut was used to select $\pi^0$'s, 
%other background reactions like $\Delta$ photoproduction and $\eta$ 
%photoproduction are not contaminated in the selected $\pi^0$ events. 
% because these thresholds are higher than the cut boundaries.   
%The backgrounds of  $\gamma p \rightarrow p\gamma$ reaction (proton
%Compton scattering) are supposed to be negligible.  
The background shape of the $\gamma p \rightarrow \pi\pi p$ reaction was 
estimated along with a 3-body phase space using a Monte Carlo simulation 
for each bin of the photon energies and the scattering angles. 
Overall factors for the background spectra were determined to minimize the 
reduced $\chi^2$ with a template fit for the experimental data.   
%The backgrounds were subtracted  
%for each bin of the photon energy and the scattering angle. 
The systematic uncertainty of the background subtraction for $\pi^0$ yields 
is 5\%. 

The spectrometer acceptance, including the efficiency for detectors and 
track reconstruction, was estimated using a Monte Carlo simulation 
based on the GEANT3 package. The acceptance, which depended on the photon 
energy and the scattering angle, was calculated.    
The systematic uncertainties of the target thickness, due to fluctuations 
of the temperature and pressure of the liquid hydrogen, was estimated to be 
1.0\%. The systematic error of the photon number normalization was 3.0\%. 
The systematic uncertainty of the aerogel \v{C}erenkov counter (AC) due to 
accidental vetoes and $\delta$-rays was measured to be lower than 
1.6\%. 

% Results & Discussion

%%% differential cross section

\begin{figure*}[h]
\begin{center}
\hspace*{-.cm}
\includegraphics*[scale=0.57]{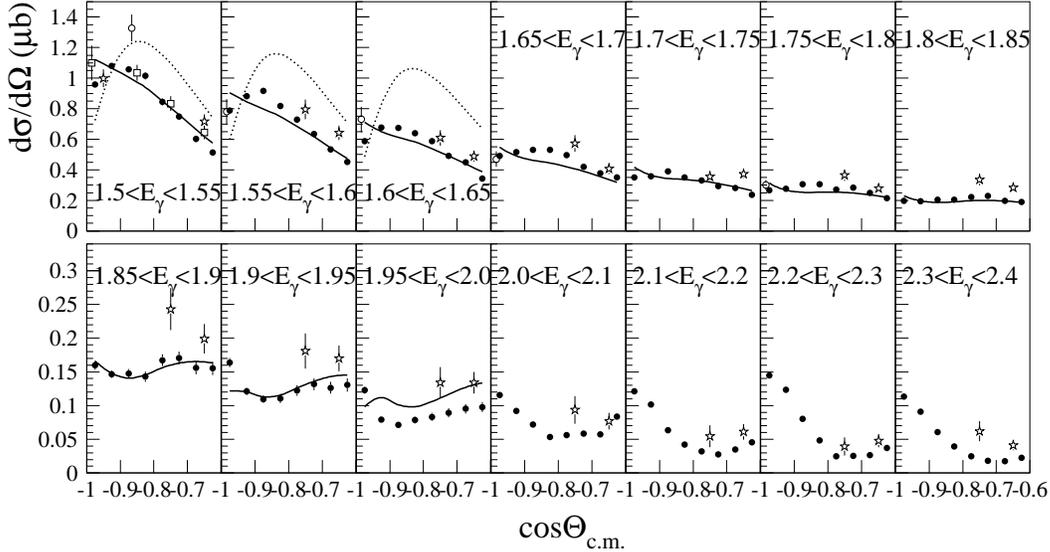}
\vspace*{-.cm}
\caption{Differential cross sections as a function of the $\pi^0$ scattering 
angle, cos$\Theta_{c.m.}$. The closed circles are the results of the present 
analysis. The open squares, open stars and open circles are the GRAAL 
data~\cite{dsg-graal},  the ELSA data~\cite{dsg-elsa}, and the Bonn data 
in 1979~\cite{dsg-bonn}, respectively.  
%~\cite{dsg-graal,dsg-elsa,dsg-spec,dsg-bonn,dsg-corn,dsg-sig-yere}.  
The solid and dotted curves are the results of the SAID~\cite{said} and the 
MAID2005~\cite{maid}, respectively. }
\label{dsgth}
\end{center}
\end{figure*}

Fig.~\ref{dsgth} shows differential cross sections as a function of the 
$\pi^0$ scattering angle. 
% in the low energy region, but does not agree with the ELSA data. 
%The $\pi^0$ photoproduction data are obtained for the 
%first time at very backward angles with high photon energies, $E_\gamma >$ 
%1.8 GeV. 
Only statistical errors are plotted in the figures 
and most of them are smaller than the size of symbols plotted. 
The LEPS data mostly agree with previously published data and 
have been measured with better statistics than the previous 
data~\cite{dsg-graal,dsg-elsa,dsg-bonn}. 

The angular distribution of the differential cross section changes around 
$E_\gamma = $ 1.8~GeV. The data show a backward peaking at cos$\Theta_{c.m.} 
<$ --0.85 above 1.8~GeV, where the $u$-channel contribution is expected to 
become large, while the data do not show the backward peaking below 1.8~GeV. 
The backward peaking suggests that the $u$-channel contribution 
is not negligibly small. The $u$-channel nucleon exchange process could 
produce the enhancement observed at backward angles and at high energies. 

The SAID (solid curves) and the MAID2005 (dotted curves) are plotted 
up to $E_\gamma =$ 2~GeV and $E_\gamma$ = 1.65~GeV,  
respectively. The LEPS data mostly agree with the SAID results. 
A small enhancement 
structure appears around cos$\Theta_{c.m.} =$ --0.7 at $E_\gamma = $ 
1.85--1.95~GeV in 
the LEPS data which have not been clearly observed in the previous data 
due to the poor statistics. The enhancement is well reproduced by the SAID 
analysis. 
%The bump structure is reproduced by the SAID. However the discrepancy with 
%the LEPS data is large at the very backward angles, 
%cos$\Theta_{c.m.} =$ --0.975 as shown in Fig~\ref{dsgth}. 
However, the SAID calculations do not reproduce the backward peaking at 
$E_\gamma = $ 1.9--2.0~GeV. 
The MAID2005 overestimates the data at cos$\Theta_{c.m.} >$ --0.9 and 
underestimates the data at the most backward angles. The 
discrepancy becomes larger as the photon energy goes higher. 
% where there is the model uncertainty of non-well-known resonances. 
These disagreements with the LEPS data are due to the lack of the precise 
data at backward angles. The partial wave analyses will be improved at higher 
energies and backward angles by including the current data, and can 
determine the property of higher-mass resonances more precisely. 

%% s-dependence of dsigma/du.

\begin{figure*}[h]
\begin{center}
\hspace*{-.cm}
\includegraphics*[scale=0.55]{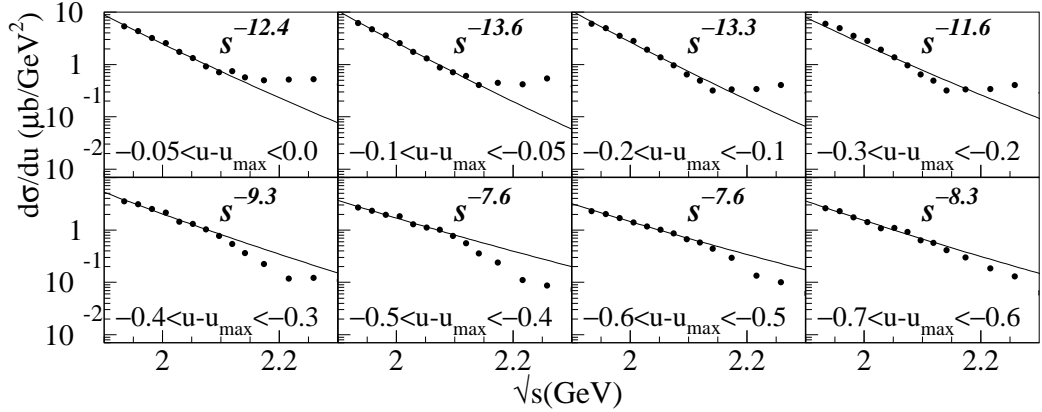}
\vspace*{-.cm}
\caption{Differential cross sections as a function of the total energy, 
$\sqrt s $, for the $\gamma p \rightarrow \pi^0 p$ reaction. The closed 
circles are the results of the present analysis. 
%The open plots are the data from other facilities, {\tiny $\Box$}~\cite{dsg-graal}, 
%$\diamond$~\cite{dsg-elsa}, {\tiny $\triangle$}~\cite{dsg-spec}, 
%$\circ$\cite{dsg-bonn}, $\ast$~\cite{dsg-corn}, $\star$~\cite{dsg-sig-yere}. 
%The thick solid and dotted curves are the results 
%of the SAID~\cite{said} and the MAID2005~\cite{maid}, respectively. 
The curves are the fitting results of $A s^{-x}$ for the data at $\sqrt s 
< $2.1~GeV, where A and x are fitting parameters~\cite{OmegaInU}.  
%The dashed curves are calculations of the model with the Born term. 
}
\label{dsgeg}
\end{center}
\end{figure*}

Fig.~\ref{dsgeg} shows differential cross sections as a function of the total 
energy, $\sqrt s $ at eight bins of $u-u_{\mbox{\tiny{max}}}$. The maximum value, 
$u_{\mbox{\tiny{max}}}$, occurs when the proton goes forward at 0$^\circ$ from the photon 
beam direction.  
The energy-dependent slope 
for $\pi^0$ production has been determined for the first time in this energy 
region.  
% for the first time in this kinematical region. 
The curves are the fitting results of $A s^{-x}$ for the data from $\sqrt s =$ 
1.93~GeV (the lowest energy) to 2.1~GeV. The slope values obtained from 
the fitting are indicated in each plot. The fitting was performed with the 
data in the range of $\sqrt s$ = 1.93--2.07=GeV and 1.93--2.20~GeV.  The 
fluctuation of the results of the slope is $\pm$ 0.5. 
 %The slope of --12 and --8 corresponds to the nucleon pole $\alpha(u) 
%= -5 $ and --3 with the Regge model,  $s^{2\alpha(u)-2}$.  

The slope becomes smaller at the larger $\mid u-u_{\mbox{\tiny{max}}}\mid$.  
If the scaling starts in this photon energy region, the data should follow 
the counting rule, $s^{2-n}$. The quantity $n$ is the total number of 
interacting photon and quarks. The value $n$ is 9 for $\pi^0$ photoproduction. 
The data becomes closer to the scaling behavior, $s^{-7}$, at the larger 
$\mid u-u_{\mbox{\tiny{max}}}\mid$ where the momentum transfer is large.  
%The data do not show this scaling behavior, yet. perhaps because the energy  
%is still too low.    

The differential cross sections sharply decrease as $E_\gamma$ goes higher, 
below 2.1~GeV.  However, above 2.1~GeV, the cross sections do not follow 
the slope determined by the data at lower energies. At $\mid u-u_{\mbox{\tiny{max}}}\mid 
< $ 0.3 (top plots in Fig.~\ref{dsgeg}), the cross sections are nearly flat 
and a small enhancement is seen at $\sqrt s > $2.1~GeV. 
%This structure is not expected from models which reproduce 
%the low energy data. 
At $\mid u-u_{\mbox{\tiny{max}}}\mid > $ 0.3 (bottom plots in Fig.~\ref{dsgeg}), the 
differential cross sections at high energies are smaller than the lines. 
The behavior of the energy distribution above $\sqrt s = $2.1~GeV 
is different from those at lower energies. 
In order to explain the current data in the $u$ channel kinematics, 
new mechanism is needed at high energies. 

%photon asymmetry
The photon beam asymmetry is sensitive to an interference of different 
diagrams. The presence of resonances can be studied by the photon beam 
asymmetry combining with cross section data. The photon beam asymmetry 
has been measured in the same kinematical 
region as the cross section measurement. By using vertically and horizontally 
polarized photons, the photon beam asymmetry is given as follows:
\begin{eqnarray}
P_\gamma\Sigma \cos2 \Phi = \frac{n N_{v}-N_{h}}{n N_{v}+N_{h}},
\label{eq:asym}
\end{eqnarray}
where $N_{v}$($N_{h}$) is the $\pi^0$ photoproduction yield with vertically
(horizontally) polarized photons, $P_\gamma$ is the photon polarization, 
$\Sigma$ is the photon beam asymmetry, and $\Phi$ is the $\pi^0$ 
azimuthal production angle which is defined using the reaction plane and 
the horizontal plane. $n$ is the normalization factor
for $N_{v}$, determined by using the numbers of horizontally polarized 
photons, $n_{h}$, and vertically polarized photons, $n_{v}$, at the target 
as $n = n_{h}/n_{v}$. The value of $n$ is $0.923$ in the present experimental 
data. The $\Phi$ dependence of the ratio $(n N_{v}-N_{h})/(n N_{v}+N_{h})$ 
was fitted with a function cos$2\Phi$ and the amplitude $P_{\gamma}\Sigma$ 
was obtained. After $P_{\gamma}$ was calculated using the photon energy 
$E_\gamma$ for Compton scattering and the laser polarization, the photon beam 
asymmetry $\Sigma$ was obtained~\cite{PRCK}. 

The correction for the background contamination from the start counter 
(SC) has been performed by using the contamination rate ($C_{SC}$) and photon 
beam asymmetries of the SC events ($\Sigma_{SC}$). The measured photon 
beam asymmetry is written as $\Sigma^{\mbox{\tiny{meas}}} = (1 - C_{bg}) \Sigma_{\pi^0} + 
C_{bg} \Sigma_{bg}$. 
The photon beam asymmetry of the pure $\gamma p \rightarrow \pi^0 p$ events,  
$\Sigma_{\pi^0}$, was obtained by determining the contamination rate 
($C_{bg}$) and the photon beam asymmetry of the events from the SC 
($\Sigma_{bg}$). 
The maximum correction was $\delta\Sigma = -0.047$ at $E_\gamma$ = 1.65~GeV 
and cos$\Theta_{c.m.}=-0.975$.  

In this analysis, a tight missing mass cut of --0.1 $<$ MM$^2$ $<$ 0.05 was 
applied for the $\pi^0$ selection in order to reduce the contamination from 
two-pion photoproduction. The background events due to the $\gamma p \rightarrow 
\pi\pi p$ reaction were corrected for in the same way as the SC 
correction. The maximum contamination rate was 6.5\%.  
The correction effect was quite small and within the statistical error.  

\begin{figure*}[h]
\begin{center}
\hspace*{-.cm}
\includegraphics*[scale=0.7]{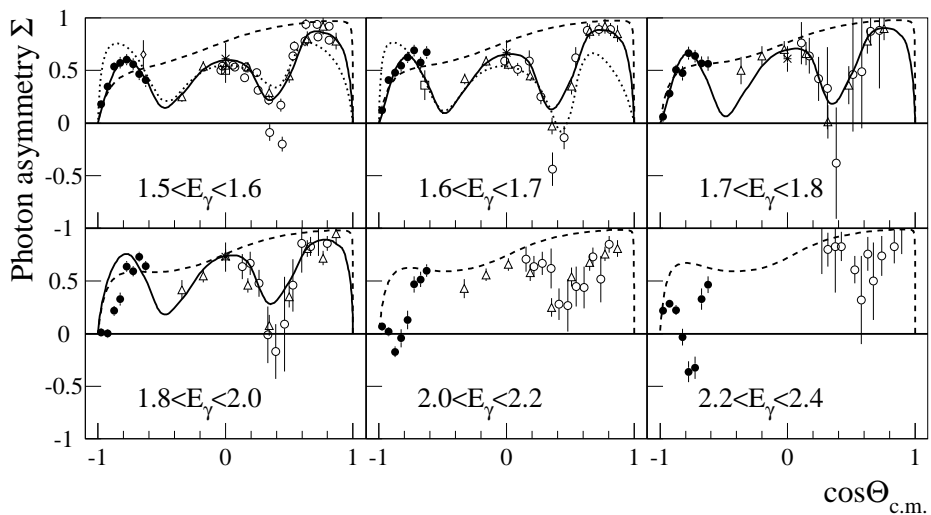}
\vspace*{-.cm}
\caption{Photon beam asymmetries $\Sigma$ as a function of the $\pi^0$ scattering 
angle, cos$\Theta_{c.m.}$. The closed circles are the results of the present 
analysis. The other plots are the data from other facilities, 
{\tiny $\Box$}~\cite{dsg-sig-yere}, {\tiny $\triangle$}~\cite{sig-dnpl2}, 
$\circ$\cite{sig-dnpl1}, $\ast$~\cite{sig-cea}. 
%~\cite{dsg-sig-yere,sig-cea,sig-dnpl1,sig-dnpl2}. 
The solid and dotted curves are the SAID results~\cite{said} 
and the MAID2005 results~\cite{maid}, respectively. The dashed curves are the 
calculations of the Born (non-resonant) term by the SL model~\cite{SL}.}
\label{asymth}
\end{center}
\end{figure*}
Fig.~\ref{asymth} shows the photon beam asymmetries. The photon asymmetries 
measured in this analysis are positive and indicate a bump structure around 
cos$\Theta_{c.m.} = -0.8$ at $E_\gamma < $ 1.8~GeV. Two bumps are observed around 
cos$\Theta_{c.m.}$ = 0 and 0.75 in the data from other facilities. A similar 
angular distribution of the photon asymmetries has been obtained at $E_\gamma 
< $ 1.5~GeV by the GRAAL collaboration~\cite{dsg-graal}.  
The MAID2005 including 13 resonances reproduces these bump structures 
although it slightly overestimates the LEPS data. The SAID agrees with 
the photon beam asymmetry data, and reproduces the bump structure up to 
$E_\gamma = $ 1.8~GeV, even though the LEPS data are not used for a fit 
in the analysis. Therefore, below 1.8~GeV, the LEPS data are explained 
by the well-established partial wave analyses with well-known resonances.  

The angular distribution changes at $E_\gamma = $ 1.8~GeV as well as the 
differential cross sections. 
A strong angular dependence appears at $E_\gamma > $ 2.0~GeV. 
The photon asymmetries show a dip structure around cos$\Theta_{c.m.}$ = --0.8. 
The data drop to a negative sign, and then rise up to a positive sign. 
%This behavior is also seen at $E_\gamma = $ 1.9 GeV although it is not so 
%significant. 
The discrepancy between the LEPS data and the SAID results is large at $E_\gamma = $ 
1.8--1.9~GeV. Here, the LEPS data will help to give a constrain for 
future partial wave analyses. 
The SL model with the Born terms only shows a positive sign and does not 
reproduce the dip structure. To explain this strong angular dependence, 
new mechanism or new resonances are required. There are some candidates 
of the resonances above 2.0~GeV and the $s$-channel contribution is still 
large at cos$\Theta_{c.m.}$ = --0.8. 

%an interference of Both the $u$-channel diagram and high-mass resonances are

A new mechanism is needed to explain both the backward enhancement in the 
cross sections and the strong angular dependence in the photon asymmetries 
observed in the LEPS data. The backward enhancement in the cross sections 
is likely produced by the $u$-channel diagrams. A new resonance is needed 
to make a strong angular dependence of the photon beam asymmetries. 
By combining both the cross section and photon asymmetry data, $\pi^0$ 
photoproduction can now be studied at backward angles and above 2~GeV. 
Recently, the applicability of SAID was extended to 3~GeV, and the proprieties of 
resonances were re-fitted by including precise data mainly covering the middle angles from 
the CLAS collaboration~\cite{CLASpi0}. The LEPS data will be useful to improve 
the analysis at backward angles.  
%The Regge model is useful to study the non-resonant term and extract signals 
%of resonances from the experimental data. Charged pion photoproduction above 
%2 GeV was studied with t-channel exchanged meson poles. They found possible 
%signals of excited baryons~\cite{Regge}. At backward angles, the data will 
%be studied by the u-channel nucleon Regge pole. 

% Summary
In summary, the $\gamma p \rightarrow \pi^0 p$ reaction has been studied 
by using linearly polarized photons at the SPring-8/LEPS facility with 
1.5--2.4~GeV. The differential cross sections have been obtained with better 
statistics than the previous data. The data at very backward angles, --1 $<$ 
cos$\Theta_{c.m.} <$ --0.9, and at $E_\gamma >$ 2.0~GeV show a backward 
peaking. It is suggested that the $u$-channel contribution is important in 
the backward region. 
In addition, photon beam asymmetries have been obtained. 
%The angular distributions  change at $E_\gamma = $ 1.8 GeV. 
Above 1.8~GeV, a dip structure is found around cos$\Theta_{c.m.}$ = --0.8 
and cannot be reproduced by using the existing models. In order to explain 
this  
structure, the presence of new high-mass resonances combined with the 
$u$-channel diagrams is required. 

%acknowledgments
We thank the staff at SPring-8 for providing excellent experimental 
conditions during the long experiment. 
This research was supported in part by the Ministry of Education, 
Science, Sports and Culture of Japan, by the National Science Council 
of the Republic of China (Taiwan), and by the National Science Foundation 
(USA).

\end{document}